\documentclass[a4paper,fleqn,useAMS,usenatbib]{mnras}

\pdfoutput=1

\usepackage{mathptmx}

\usepackage[T1]{fontenc}
\usepackage{ae,aecompl}

\usepackage[dvips]{graphicx}
\usepackage{longtable}
\usepackage{amsmath}
\usepackage{amsfonts}
\usepackage{amssymb}
\usepackage{relsize}
\usepackage{tablefootnote}
\usepackage{comment}
\usepackage{booktabs}
\usepackage[dvipsnames]{xcolor}
\usepackage[normalem]{ulem}
\usepackage{multirow}
\usepackage{wasysym}
\usepackage[%
        ]{hyperref}
        
\hypersetup{
	colorlinks=true,	
	urlcolor=MidnightBlue,		
	pdfpagelabels=true,
	hypertexnames=true,
	plainpages=false,
	naturalnames=true,
	pdftitle={Phase curve and occultation of WASP-100b},    
	pdfauthor={Jansen \& Kipping},     
	linkcolor=WildStrawberry,          
	citecolor=ForestGreen,        
}

\newcommand{\Kepler}{{\it Kepler}}
\newcommand{\TESS}{{\it TESS}}
\newcommand{\phasma}{{\tt phasma}}

\newcommand{\multi}{{\sc MultiNest}}
\newcommand{\jup}{\jupiter}

\title[Phase curve and occultation of WASP-100b]{
Detection of the phase curve and occultation of WASP-100b with \TESS
}
\author[Jansen \& Kipping]{Tiffany Jansen$^{1}$\thanks{E-mail:
\href{mailto:jansent@astro.columbia.edu}{jansent@astro.columbia.edu}} and David Kipping$^{1,2}$\\
$^{1}$Dept. of Astronomy, Columbia University, 550 W 120th Street, New York NY 10027\\
$^{2}$Flatiron Institute, 162 5th Av., New York, NY 10010}

\date{Accepted by MNRAS March 19, 2020}

\pubyear{2020}

\begin{document}
\label{firstpage}
\pagerange{\pageref{firstpage}--\pageref{lastpage}}
\maketitle

\begin{abstract}
We report the detection of the full orbital phase curve and occultation of the
hot-Jupiter WASP-100b using \TESS\ photometry. The phase curve is isolated
by suppressing low frequency stellar and instrumental modes using both a
non-parametric harmonic notch filter (\phasma) and semi-sector long
polynomials. This yields a phase curve signal of $(73\pm9)$\,ppm
amplitude, preferred over a null-model by $\Delta\mathrm{BIC}=25$, indicating very
strong evidence for an observed effect. We recover the occultation event with
a suite of five temporally localized tools, including Gaussian processes
and cosine filtering. This allows us to infer an occultation depth of
$(100\pm14)$\,ppm, with an additional $\pm16$\,ppm systematic error from the
differences between methods. We regress a model including atmospheric
reflection, emission, ellipsoidal variations and Doppler beaming to the
combined phase curve and occultation data. This allows us to infer that
WASP-100b has a geometric albedo of $A_g = 0.16^{+0.04}_{-0.03}$ in the \TESS\
bandpass, with a maximum dayside brightness temperature of $(2710\pm100)$\,K and a
warm nightside temperature of $(2380^{+170}_{-200})$\,K. Additionally, we find evidence
that WASP-100b has a high thermal redistribution efficiency, manifesting as a
substantial eastward hotspot offset of $(71^{+2}_{-4})^{\circ}$. These results
present the first measurement of a thermal phase shift among the phase curves
observed by \TESS\ so far, and challenge the predicted efficiency of heat
transport in the atmospheres of ultra-hot Jupiters.
\end{abstract}


\section{Introduction}
\label{sec:intro}

Hidden within the light curve of a star hosting an exoplanet is the light from the planet itself, waxing and waning as it traverses its orbit, reflecting and re-radiating its star's incident rays. Upon folding a light curve into a function of orbital phase, flux modulations due to the presence of an orbiting companion can become prominent above the noise. In addition to the waxing and waning of the planet's atmospheric phase curve, there are sinusoidal signatures caused by the apparent change in surface area of the tidally distorted star and the Doppler beaming of its radiation as the star orbits its system's center of mass. While these stellar signals are indicative of the star-planet mass ratio, the amplitude and symmetry of the planetary phase curve reveal important atmospheric characteristics of the planet in question, such as its albedo, thermal redistribution efficiency, day-to-night temperature contrast, or whether a significant atmosphere exists at all (e.g. \citealt{knutson:2007, hu:2015, kreidberg:2019, parmentier:2018a} and references therein). 

Due to their often bloated radii and proximity to their host stars, hot-Jupiters are excellent candidates for atmospheric characterization. WASP-100b is one such hot-Jupiter discovered transiting a ${\sim}6900$\,K F2 star with a radius of $R_P = (1.69\pm0.29)\,R_{\jup}$ \citep{hellier:2014}. Observations of this system with the Euler/CORALIE spectrograph reveal an eccentricity consistent with zero, and together with the transit data give a mass of $M_P = (2.03\pm0.12)\,M_{\jup}$ \citep{hellier:2014}. With an orbital period of 2.9\,days and a semi-major axis of $a = 0.046$\,AU reported by \cite{hellier:2014}, WASP-100b is likely tidally locked in a synchronous orbit to its star \citep{guillot:1996}.

At the time of writing, the Transiting Exoplanet Survey Satellite (\TESS) has completed the Southern-hemispherical half of its primary mission to survey the brightest stars for transiting exoplanets \citep{ricker:2015}, and is well into its second half of the mission to survey the northern hemisphere. Only a handful of full phase curves have been measured in the \TESS\ data prior to this study \citep{shporer:2019, daylan:2019, bourrier:2019, wong:2019}. With an orbital period of 2.9\,days, WASP-100b is the longest-period planet to have a full phase curve and occultation depth measured in the \TESS\ data to-date. This is due in part to its location in the continuous viewing zone of the \TESS\ field of view. By observing WASP-100 in each of the 13 observational sectors, we are able to increase the signal-to-noise ratio of WASP-100b's phase curve by stacking 76 of its orbits.

In this study, we present the first occultation measurements of WASP-100b and aim to constrain its atmospheric characteristics such as albedo, thermal redistribution efficiency, intrinsic thermal scaling factor, and day-to-night temperature contrast. In Section~\ref{sec:data}, we describe our data processing methods, and in Section~\ref{sec:occultation} we measure the occultation depth of WASP-100b. In Section~\ref{sec:model}, we describe the phase curve model used in our regression analysis, which is detailed in Section~\ref{sec:regression}. The constraints we are able to place on the atmospheric characteristics of WASP-100b are presented in Section~\ref{sec:results} and discussed in Section~\ref{sec:discussion}.

\section{Extracting the Phase Curve}
\label{sec:data}

\begin{figure}
    \centering
    \includegraphics[width=0.5\textwidth]{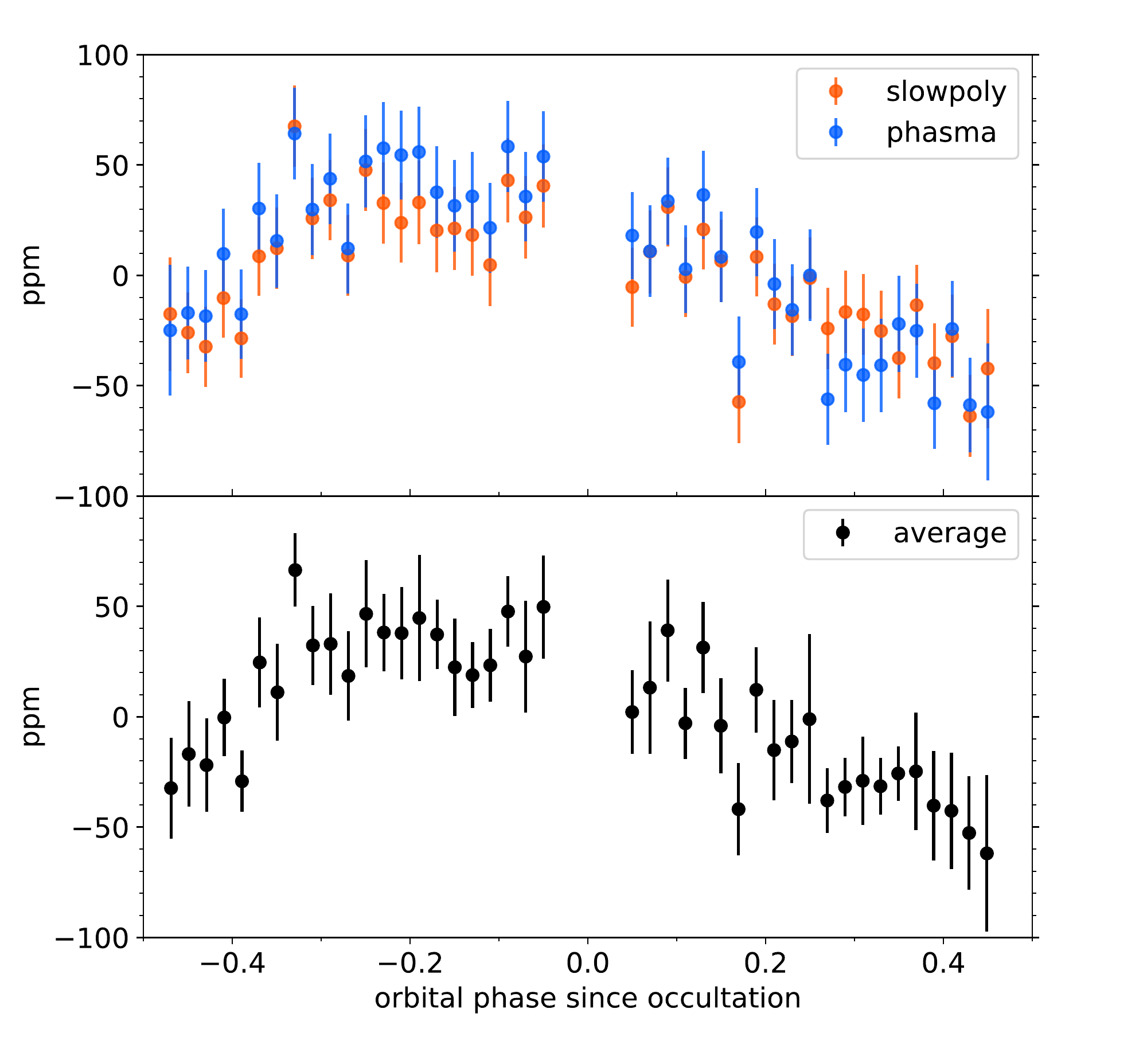}
    \caption{Phase curve of WASP-100 detrended by the polynomial method (top), the moving median method \phasma\ (middle), and the average of the two (bottom) which is used in the analysis. The occultation and transit events at $x = 0.0, \pm 0.5$ have been removed both from this figure and before the regression analysis. }
    \label{fig:phasecurve-methods}
\end{figure}

We analyze the 2-minute cadence Pre-search Data Conditioning simple aperture photometry (PDCSAP) light curves of the WASP-100 system (TIC 38846515, TOI 106) from \TESS\ Sectors 1 -- 13, downloaded from the Mikulski Archive for Space Telescopes on September 9th, 2019. All images were taken with \TESS\ Camera 4. The PDCSAP light curves have been corrected for systematics with the Science Processing Operations Center pipeline \citep{jenkins:2016}. Data with quality flags indicating any anomalous behavior were removed prior to analysis\footnote{Data quality flag descriptions can be found in Section 9 of the \TESS\  \hyperlink{https://ntrs.nasa.gov/archive/nasa/casi.ntrs.nasa.gov/20180007935.pdf}{Science Data Products Description Document}.}.

\subsection{Outlier removal}

We first remove any remaining outliers in the PDCSAP time series using a standard moving filtering approach. We evaluate a moving median smoothing function through the time series with a 10-point window, which we then linearly interpolate and evaluate the distance of the data away from this function. Points greater than $f \sigma$ away are classified as outliers, where $\sigma$ is given by $1.4826$ multiplied by the median absolute deviation of the residuals (a robust estimator of the standard deviation, \citealt{huber:1981}), and $f$ is set to 4. We choose 4-sigma on the basis that this results in an expectation that no more than one non-outlier data point will be erroneously removed, assuming Gaussian noise.

\subsection{Nuisance signal detrending}

We apply two methods for removal of long-term stellar variability and low frequency systematics. In order to correctly apply the methods described in the remainder of this section, it was necessary to first identify significant gaps of missing data in the light curve and concentrate on each continuous section of data individually for its reduction. We define a ``significant'' gap to be one which is greater than 10\% of the moving median window. For both detrending methods this necessitates separating each sector of data at \TESS' data downlink gap, which lasts on the order of $\sim 1$\,day in the middle of each sector's baseline.

For the first method, we fit a low-order polynomial function to each semi-sector of the cleaned light curve using weighted linear least squares. The idea is that the polynomial acts as a low-cut filter, but in reality polynomials can present complex behaviour in the frequency domain. For each semi-sector, we mask the transits and then evaluate the Akaike Information Criterion \citep{akaike:1974} of the polynomial fit from $1^{\mathrm{st}}$ to $20^{\mathrm{th}}$ order. The preferred model (lowest AIC score) is adopted and used to normalize that semi-sector. This polynomial treatment is a fairly standard way of removing long term trends in phase curve analysis and has been used by previous studies of both \Kepler\ and \TESS\ data (e.g. \citealt{wong:2020, shporer:2019}).

We then separately apply \phasma\footnote{The development version of \phasma\ can be downloaded at \hyperlink{https://github.com/tcjansen/phasma}{https://github.com/tcjansen/phasma}}, a non-parametric moving median algorithm that operates as a harmonic notch filter with a kernel equal to the orbital period $P$, removing nuisance signals which
are out of phase with the phase curve (e.g. long-term stellar variability and residual systematics). A mathematical description of this method can be found in Appendix~\ref{sec:appendix}, and is further described in Section~2.2 of \cite{jansen:2018}. Unlike the polynomial method, {\tt phasma} not does assume any particular functional form for the nuisance signal, which leads to a generally less-precise but more-accurate detrending \citep{jansen:2018}.

The primary transit and occultation are removed prior to the \phasma\ detrending to avoid contaminating the moving median function and regression analysis. The semi-sector light curves are then stitched and phase folded, then binned into 500 points in phase using a weighted mean (where the weights comes from the PDCSAP uncertainties). During this binning, we calculate new uncertainties for the binned points directly from the standard deviation of the data within that phase bin. In this way the errors are empirically derived. As apparent from Figure~\ref{fig:phasecurve-methods}, the two methods produce very similar phase curves which provides confidence that the reconstructions are not purely an artefact of the algorithms used. For our regression analysis, we take the mean of the binned {\tt phasma} detrended phase curve and the binned polynomial detrended phase curve to obtain the data which are modeled in Section~\ref{sec:model}.

\subsection{Background contamination}

According to The Exoplanet Follow-up Observing Program (ExoFOP) for \TESS\, there are 10 other sources within 1 arcminute of WASP-100, the brightest at a separation of 28.1 arcseconds and about 13 times fainter than our target. With the \TESS\ pixel width of 21 arcseconds, this source lies in an adjacent pixel to WASP-100. Additionally, WASP-100 shares its central pixel with another object at a separation of 3.78 arcseconds, which is about 400 times fainter than our target.

We correct for aperture contamination by background sources such as these by using these blend factors and following the prescription of \cite{kipping:2010}. Blend factors are obtained from the \TESS\ crowding metric ``CROWDSAP'', which is defined to be the ratio of the flux of the target to the total flux in the aperture\footnote{CROWDSAP definition from the \TESS\ \hyperlink{https://ntrs.nasa.gov/archive/nasa/casi.ntrs.nasa.gov/20180007935.pdf}{Science Data Products Description
Document}.}. The average crowding metric across all 13 sectors in which WASP-100 is observed is $0.93 \pm 0.01$ (i.e. a contamination of 7\% in flux).

\section{Occultation}
\label{sec:occultation}

\begin{figure*}
    \centering
    \includegraphics[width=0.65\textwidth]{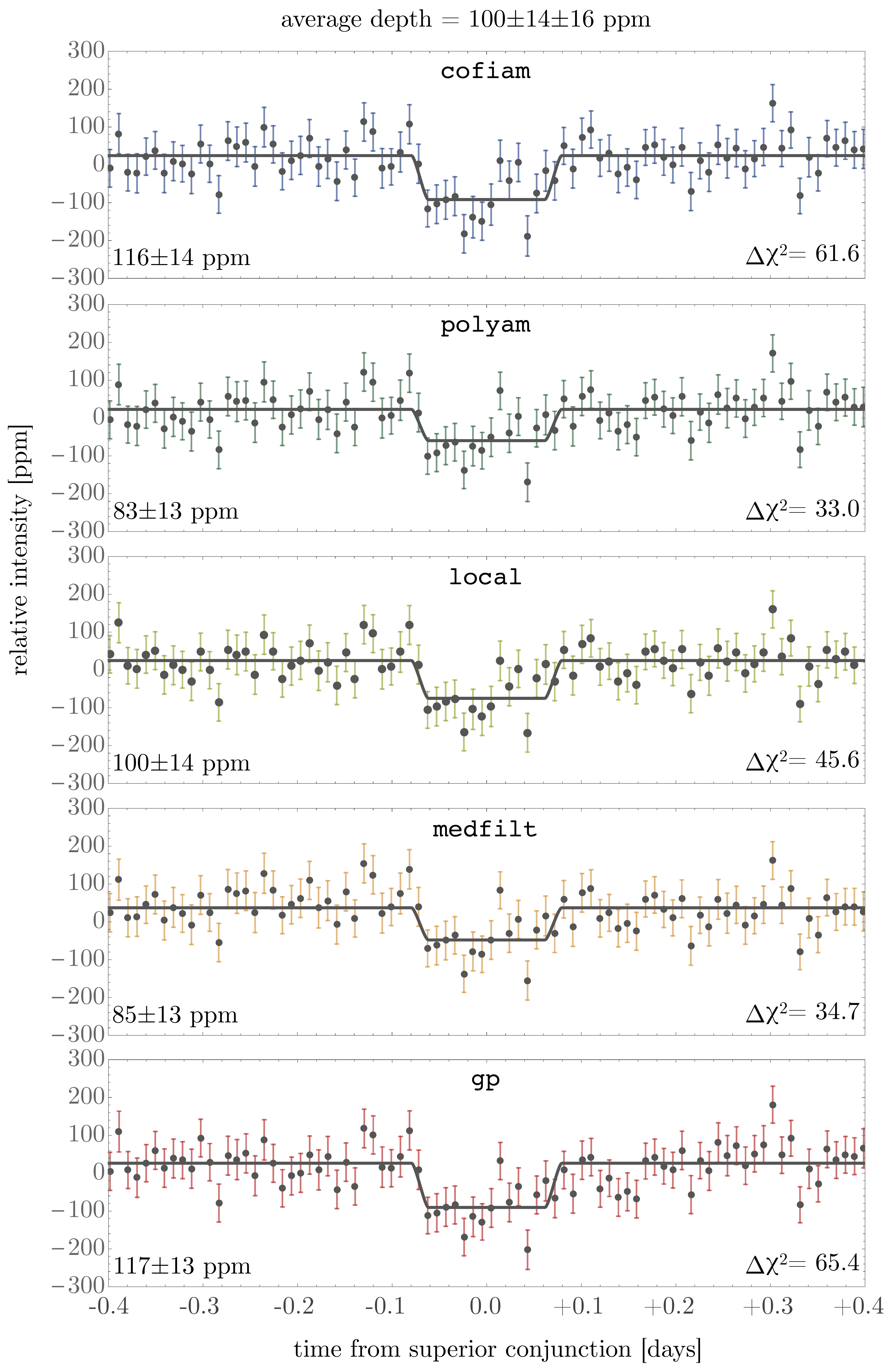}
    \caption{Comparison of the five different methods used to detrend the WASP-100b occultation events observed by \TESS. We use the variability between the different methods to assign a $\pm16$\,ppm ``systematic'' error
    on top of the $\pm14$\,ppm random error.}
    \label{fig:occultation}
\end{figure*}

Formally, the \phasma\ algorithm is not optimized for sharp features such as transits and occultations. This is because the convolution of a transit's Fourier transform (characterized by harmonics of the transit duration (\citealt{waldmann:2012})), with \phasma's harmonic notch filter (characterized by finite width notches) will, in general, lead to bleeding of the transit's spectral power out of the notches, thereby distorting the transit profile \citep{jansen:2018}. For this reason, we elect to detrend the occultation data using a distinct approach from \phasma.

Specifically, we follow the approach of \cite{teachey:2018} who detrend the photometry with a multitude of common algorithms to ensure the result is robust against detrending choices. We used {\tt CoFiAM} \citep{kipping:2013a}, BIC-guided polynomial detrending to semi-sectors, BIC-guided polymonial detrending to local occultation regions, median filtering, and a Gaussian process. We choose local occultation regions spanning $\pm2.5$ transit durations from the time of occultation such that the baseline is larger than the occultation window, but small enough to exclude a phase curve signature. After detrending the occultations, the signals were coherently phase-folded. The phase folded occultation resulting from each detrending method can be seen in Figure \ref{fig:occultation}.

We find clear evidence for an occultation event at the expected location for a near-circular orbit for all five methods. The average depth of the occultation event is $(100\pm14)$\,ppm with an additional systematic error of $\pm16$\,ppm originating from the differences between the methods. The depth was obtained by regressing a \cite{mandel:2002} transit model to the transit light curves, and then scaling that best fitting template light curve to the occultation event (with limb darkening turned off when applied to the occultation).

This formally assumes a circular orbit with a single free parameter describing the ratio of the transit-to-occultation depth ratio. To check that the orbital eccentricity is indeed consistent with a circular orbit, we allow the time of occultation to vary while fitting and measure a median offset of $t_{occ} = -120\pm329$\,s, with an additional systematic error of $\pm 69$\,s across the five methods. This yields an upper limit on the eccentricity of  $|e\cos{\omega}| < 0.0029$ to $3\sigma$ confidence  (see Section 4.4.1 in \citealt{kipping:2011}).

The weighted mean of the measured occultation depths is then used in our regression's likelihood function in order to constrain the parameters contributing to the thermal component of the full phase curve.

\section{Out-of-Transit Phase Curve Model}
\label{sec:model}

We model the out-of-transit phase curve of the WASP-100 system as a sum of the planet's atmospheric phase curve, photometric effects in-phase with the orbital period by the stellar host, and a constant term $\gamma$ which accounts for possible residual noise from the detrending process,  

\begin{equation}
\mathcal{F} = F_P(\phi, A_B, f, \epsilon) + F_{\star}(\phi, A_{\mathrm{beam}}, A_{\mathrm{ellip}}) + \gamma.
\label{eq:forwardmodel}
\end{equation}

This expression gives the flux of WASP-100 normalized by the average flux of the star as a function of orbital phase $\phi$. Here, we define the orbital phase as $\phi \equiv 2\pi\left(P^{-1}[t - t_0] + \frac{1}{2}\right)$, where $t_0$ is the transit ephemeris and $P$ is the orbital period. Note that this expression is shifted from the canonical definition of orbital phase by $\pi/2$ such that the transit occurs at $\phi = \pm \pi$ and the occultation occurs at $\phi=0$. This is simply to maintain consistency with the model described in the remainder of this section.

The atmospheric contribution of the phase curve is described by a sum of the thermal component and the reflective component,

\begin{equation}
F_P(\phi, A_B, f, \epsilon) = F_T(\phi, A_B, f, \epsilon) + F_R(\phi, A_B).
\end{equation}

We model the thermal component $F_T(\phi, A_B, f, \epsilon)$ with the Bond albedo $A_B$, a thermal redistribution efficiency factor $\epsilon$ (defined as in \citealt{cowan:2011}), and an intrinsic thermal scaling factor $f$. The thermal redistribution efficiency is here defined to be the ratio between the radiative timescale of the planet's photosphere and the difference between the frequencies at which the photosphere rotates about the planet and the surface rotates about its axis. In other words, if the atmospheric mass heated at the substellar point is redistributed about the surface much faster than the heat gets reradiated, the planet would be described as having a large redistribution efficiency $\epsilon$, typically $\epsilon\gg1$. Conversely, a planet with relatively no heat redistribution would be described as having $\epsilon$ = 0. For a planet which has winds moving in a direction opposite of the planetary rotation, $\epsilon$ is defined to be negative. The intrinsic thermal factor $f$ is simply a temperature scaling factor which accounts for any deviation from the equilibrium temperature due to e.g. the presence of greenhouse gases in the atmosphere or interior heat from a dynamic core. 

We express the thermal emission component of the phase curve as

\begin{equation}
\begin{split}
F_T(\phi, A_B, f, \epsilon) = &\frac{1}{\pi B_{\tau,\star}}\left(\frac{R_P}{R_{\star}}\right)^{2}\\
&\times\int_{-\frac{\pi}{2}}^{\frac{\pi}{2}}\int_{-\frac{\pi}{2}}^{\frac{\pi}{2}}B_{\tau,P}[T(\phi, \theta, \Phi)]\cos^{2}{\theta}\cos{\Phi}d\theta d\Phi,
\end{split}
\end{equation}

where $B_{\tau,\star}$ is the Planck function of the host star convolved with the wavelength response function of \TESS\footnote{Approximately 600 - 1000 nm}, $R_P$ is the radius of the planet, $R_{\star}$ the radius of the star, and $B_{\tau,P}[T(\phi,\theta,\Phi)]$ is the temperature distribution dependent blackbody curve of the planet convolved with the \TESS\ bandpass,

\begin{equation}
\begin{split}
B_{\tau,P}[T(\phi,\theta,\Phi)] = &\int_{\lambda}\tau_{\lambda}\frac{2hc^{2}}{\lambda^{5}}\\
&\times\left[\exp{\left(\frac{hc}{\lambda k_{B}}\frac{1}{ T(\phi,\theta,\Phi)}\right)} - 1\right]^{-1}d\lambda
\end{split}
\label{eq:planck}
\end{equation}

where $\tau_{\lambda}$ is the response function of \TESS, and $T(\phi,\theta,\Phi)$ is the phase-dependent temperature distribution across the planet's surface, where $\phi$, $\theta$ and $\Phi$ represent the orbital phase and planetary latitude and longitude as viewed in the observer's frame of reference, respectively. For our models we have chosen a surface resolution of $15^{\circ} \times 15^{\circ}$ in latitude and longitude, where further increasing the resolution only changes the thermal amplitude on the order of one-hundredth of a percent.
It should be noted that $\Phi$ and $\theta$ are independent of phase, where $\Phi \equiv 0$ in the direction of the observer.

We borrow from \cite{hu:2015} to define the phase-dependent temperature distribution $T(\phi, \theta, \Phi)$ to be equal to

\begin{equation}
T(\phi, \theta, \Phi) = f T_{0}(\theta)\mathcal{P}(\epsilon, \xi)
\label{eq:tempdist}
\end{equation}

where $T_{0}$ is the sub-stellar temperature and $\mathcal{P}$ is the thermal phase function, which for a planet on a circular orbit can be expressed by Equation~(10) in \cite{cowan:2011}:

\begin{equation}
\frac{d\mathcal{P}}{d\xi} = \frac{1}{\epsilon}(\text{max}(\cos{\xi}, 0) - \mathcal{P}^{4})
\label{eq:therm_P}
\end{equation}

where max($\cos\xi$, 0)  = $\frac{1}{2}(\cos{\xi} + |\cos{\xi}|)$, i.e. a cosine function truncated at negative values. We borrow our notation from \cite{hu:2015}, where $\xi$ represents the local planetary longitude defined for all points in phase to be $\xi \equiv \Phi - \phi$ for a synchronously rotating planet. The phase term $\phi$ ranges from $-\pi$ to $\pi$ and is defined to be zero at the occultation. For a planet with prograde rotation, $\xi = 0$ at the sub-stellar longitude, $\xi = -\pi/2$ at the dawn terminator, and $\xi = \pi/2$ at the dusk terminator.

Equation (\ref{eq:therm_P}) does not have an analytic solution, so we solve it numerically using {\tt scipy}'s ODE integrator, where we set the initial conditions equal to the approximated expression for $\mathcal{P}_{\mathrm{dawn}}$ stated in the Appendix of \cite{cowan:2011},

\begin{equation}
\mathcal{P}_{\mathrm{dawn}} \approx \left[\pi + (3\pi/\epsilon)^{4/3}\right]^{-1/4}.
\end{equation}

The sub-stellar temperature as a function of planetary latitude $\theta$ is expressed by

\begin{equation}
T_{0}(\theta) = T_{\star}\left(\frac{R_{\star}}{a}\right)^{1/2}(1 - A_{B})^{1/4}\cos{\theta}^{1/4}
\end{equation}

where $T_{\star}$ is the effective temperature of the host star and $a$ the semi-major axis.

The reflection component of the atmospheric phase curve $F_R(\phi)$ is assumed to be symmetric, and is proportional to the geometric albedo $A_{g}$,

\begin{equation}
F_R(\phi, A_B) = \left(\frac{R_P}{a}\right)^{2}\frac{2}{3}A_{B}\frac{1}{\pi}\left[\sin{|\phi|} + (\pi - |\phi|)\cos{|\phi|}\right]
\end{equation}

where we adopt the Lambertian approximation such that $A_{g} = \frac{2}{3}A_{B}$. According to \cite{seager:2000} and \cite{cahoy:2010}, this is a fine approximation under the assumption that the atmosphere is reflecting homogeneously. Caveats of this assumption and the expectation of symmetry are discussed in Section~\ref{sec:discussion}.

The second term in Equation \ref{eq:forwardmodel} describes the contribution to the phase curve by the host star,

\begin{equation}
F_{\star}(\phi, A_{\mathrm{beam}}, A_{\mathrm{ellip}}) = -A_{\mathrm{beam}}\sin(\phi) - A_{\mathrm{ellip}}\cos(2\phi).
\end{equation}

The first sinusoidal term $A_{\mathrm{beam}}\sin(\phi)$ accounts for relativistic beaming of the star's radiation as it orbits the system's center of mass \citep{rybicki:1979}. Ellipsoidal variations due to any tidal distortion of the host star by the close-in companion can be described by the second harmonic of the orbital period as $A_{\mathrm{ellip}}\cos(2\phi)$ \citep{morris:1985}. The amplitudes $A_{\mathrm{beam}}$ and $A_{\mathrm{ellip}}$ are left as free parameters in the regression and are described in further detail in Section~{\ref{sec:regression}}.

The third term in our phase curve model $\gamma$ accounts for a possible offset in the vertical alignment of our model from the data (not to be confused with the phase offset of the brightness maximum). Such an offset, which is constant in phase, could be a product of the normalization in the polynomial detrending process, an effect of \phasma's harmonic notch filter, or residual stellar noise (see Appendix~\ref{sec:appendix}).

\section{Phase Curve Regression Analysis}
\label{sec:regression}

Bayesian inference of the model parameters, conditioned upon our phase curve data, is achieved using {\tt emcee} \citep{dfm:2013}. We allow the Bond albedo $A_B$, thermal redistribution efficiency $\epsilon$, intrinsic thermal factor $f$, vertical adjustment term $\gamma$, scaled Doppler beaming amplitude $\left(A\alpha^{-1}\right)_{\mathrm{beam}}$, and scaled ellipsoidal variation amplitude $\left(A\alpha^{-1}\right)_{\mathrm{ellip}}$ to vary as free parameters. Although the scaling factors $\alpha_{\mathrm{beam}}$ and $\alpha_{\mathrm{ellip}}$ can be approximated analytically, we instead choose to leave them as free parameters to account for the uncertainty in their values. We obtain $10^6$ samples from $2.5\times10^4$ steps across 40 walkers, burning the first half of the chains for a remaining total of $5\times10^5$ samples. The chains were inspected to ensure they had converged and achieved adequate mixing.

\subsection{Transit fits}
\label{sec:transitfit}

Many of the transit parameters affect the shape of the occultation and phase
curve. For this reason, it is helpful to determine a-posteriori distributions
for the transit terms, which can then serve as informative priors in the
analysis of these effects. To this end, we detrended and regressed the \TESS\
data of WASP-100b's transits for all thirteen available sectors. Detrending was
performed by method marginalization over {\tt CoFiAM},
a Gaussian process, a moving median filter, semi-sector polynomials, and
epoch-localized polynomials. The resulting light curve was regressed using
\multi\ coupled to the \citet{mandel:2002} forward transit model. 

For these fits, we adopted uniform priors on the transit parameters, including
the quadratic limb darkening coefficients re-parameterized to the $q_1$-$q_2$
system \citep{kipping:2013b}. The only exception to this was for the stellar density,
for which we adopt a Gaussian prior of $(440\pm100)$\,kg\,m$^{-3}$, which comes
from the isochrone analysis described in the following subsection
(Section~\ref{sec:priors}).

\begin{figure}
\centering
\includegraphics[width=0.45\textwidth]{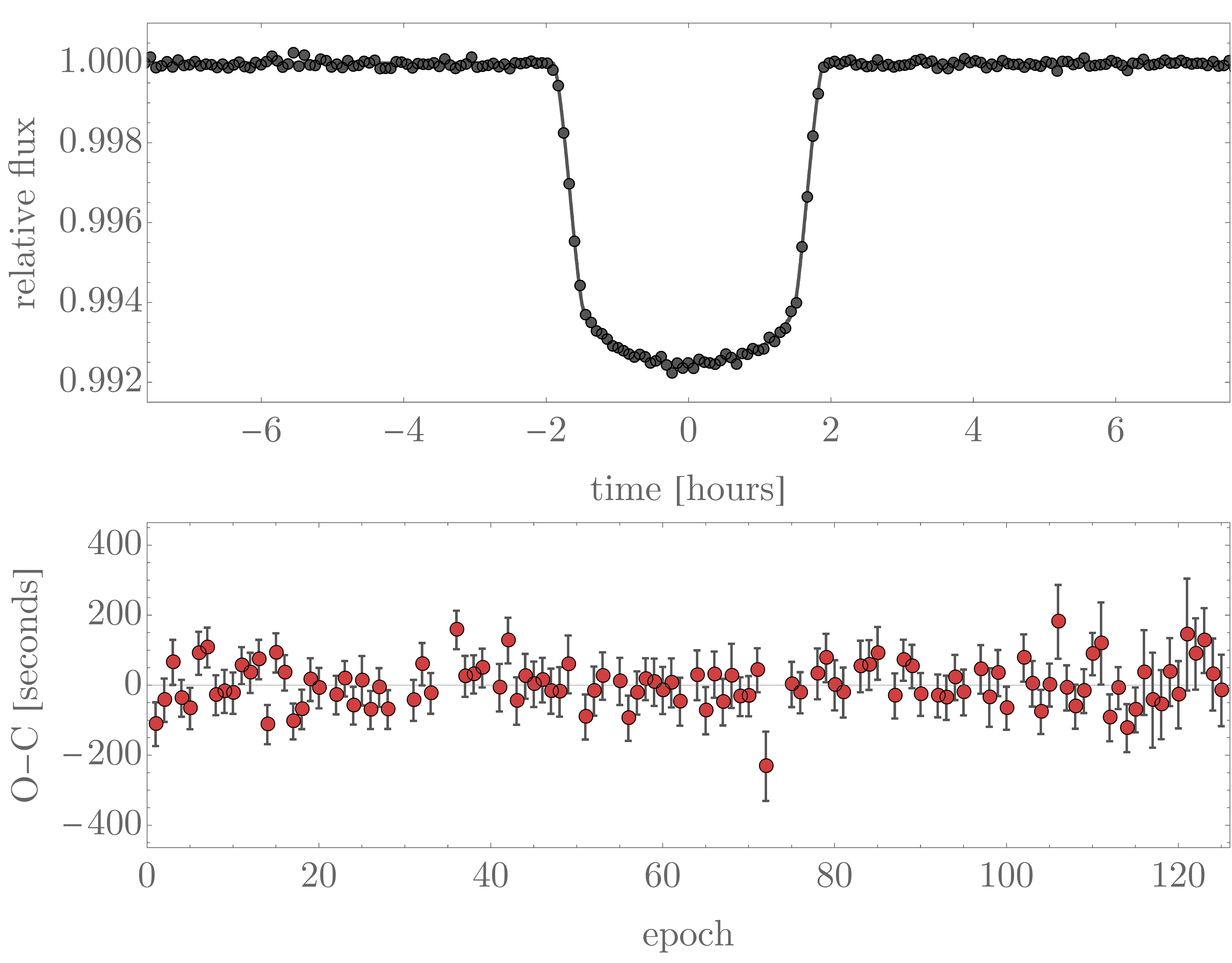}
\caption{
Top: Maximum a-posteriori light curve solution to the \TESS\ data of WASP-100b. Here we show 13 sectors of photometry folded upon the best fitting ephemeris.
Bottom: Transit timing variations of WASP-100b measured using \TESS. We find no evidence for periodicity within the observed times.
}
\label{fig:transitfit}
\end{figure}

The maximum \textit{a-posteriori} phase folded light curve model is shown in the top panel of 
Figure~\ref{fig:transitfit}, which well describes the \TESS\ data. The unbinned
residuals to this solution display a standard deviation of 1227\,ppm. The
one-sigma \textit{a-posteriori} credible intervals for the seven fitted
parameters are given in Table~\ref{tab:parameters}. 

To determine if there are any transit timing variations in the light curve of WASP-100, we fit the transits in each
sector assuming global transit shape parameters and unique
transit times. We find no
evidence for any periodicity in the transit time residuals (see the bottom panel of Figure \ref{fig:transitfit}), which have a standard deviation of
65.1 seconds. Because the standard deviation of the residuals is very close to the median formal timing
uncertainty of 64.3 seconds, we report there being no evidence for significant TTV signals.

\subsection{Prior distributions}
\label{sec:priors}

One useful piece of prior information in our analysis comes from the star itself. We elected to derive our own stellar parameter posteriors using an isochrone analysis of archival data. Specifically, we used T. Morton's {\tt isochrone} package \citep{morton:2015} with the Dartmouth stellar evolutionary models to constrain the host star's fundamental parameters. For this analysis, we used $V = 10.8 \pm 0.06$ \citep{hog:2000}, $T_{\mathrm{eff}} = 6900 \pm 120$, $\mathrm{[Fe/H]} = -0.03 \pm 0.10$, $\log g = 4.35 \pm 0.17$ \citep{hellier:2014} and the Gaia Data Release 2 parallax of $2.7153 \pm 0.0204\,$mas \citep{gaia:2018}. The resulting stellar parameters can be seen in Table~\ref{tab:parameters}.

For the atmospheric parameters $A_B$, $\epsilon$, and $f$, we sample from uniform priors spanning [0,1], [-10, 50], and [1, 5], respectively. The vertical offset $\gamma$ is sampled from a uniform prior spanning [-200, 200] ppm. We can construct more informative priors for the mass-induced amplitudes $A_{\mathrm{ellip}}$ and $A_{\mathrm{beam}}$ from the spectroscopic radial velocity measurements of WASP-100 \citep{hellier:2014} coupled with the posteriors of the transit light curve parameters and the characteristics of the host star modeled with {\tt isochrones}, which can be seen in Table~\ref{tab:parameters}. 

The scaled amplitude for the magnitude variation due to tidal distortion of the host star can be approximated as

\begin{equation}
\left(A\alpha^{-1}\right)_{\mathrm{ellip}} \approx K_{\mathrm{RV}} \left(\frac{R_{\star}}{a}\right)^{3} \frac{P}{2\pi a} \sin{i},
\label{eq:Aellip}
\end{equation}

where $K_{\mathrm{RV}}$ is the radial velocity semi-amplitude, $R_{\star}$ the radius of the star, $a$ the semimajor axis of the companion, $P$ the orbital period, and $i$ the inclination of the system in the observer's line of sight.
For this expression we used the approximations of \cite{faigler:2011} and \cite{morris:1993} on the theoretical derivations by \cite{kopal:1959} and the momentum relation $m_p \sin(i) = M_*K_{rv}P(2\pi a)^{-1} $. The scaling factor $\alpha_{\mathrm{ellip}}$ contains the limb-darkening and gravity-darkening coefficients which we do not attempt to estimate, but instead leave as a free parameter with a uniform prior spanning the estimated range for F-G-K stars of [1.0, 2.4] \citep{faigler:2011}.

The beaming amplitude can be described by

\begin{equation}
\left(A\alpha^{-1}\right)_{\mathrm{beam}} = \frac{4K_{\mathrm{RV}}}{c}.
\label{eq:Abeam}
\end{equation}

where the scaling factor $\alpha_{\mathrm{beam}}$ accounts for deviations from the beaming effect in a bolometric observation (the right side of Eq. \ref{eq:Abeam}) due to observing a spectrum that gets Doppler shifted within a finite bandpass \citep{loeb:2003, faigler:2011}. We adopt a conservative prior for $\alpha_{\mathrm{beam}}$, which we set to be uniform in the range [0.5, 1.5].

\begin{figure*}
\centering
\includegraphics[width=\textwidth]{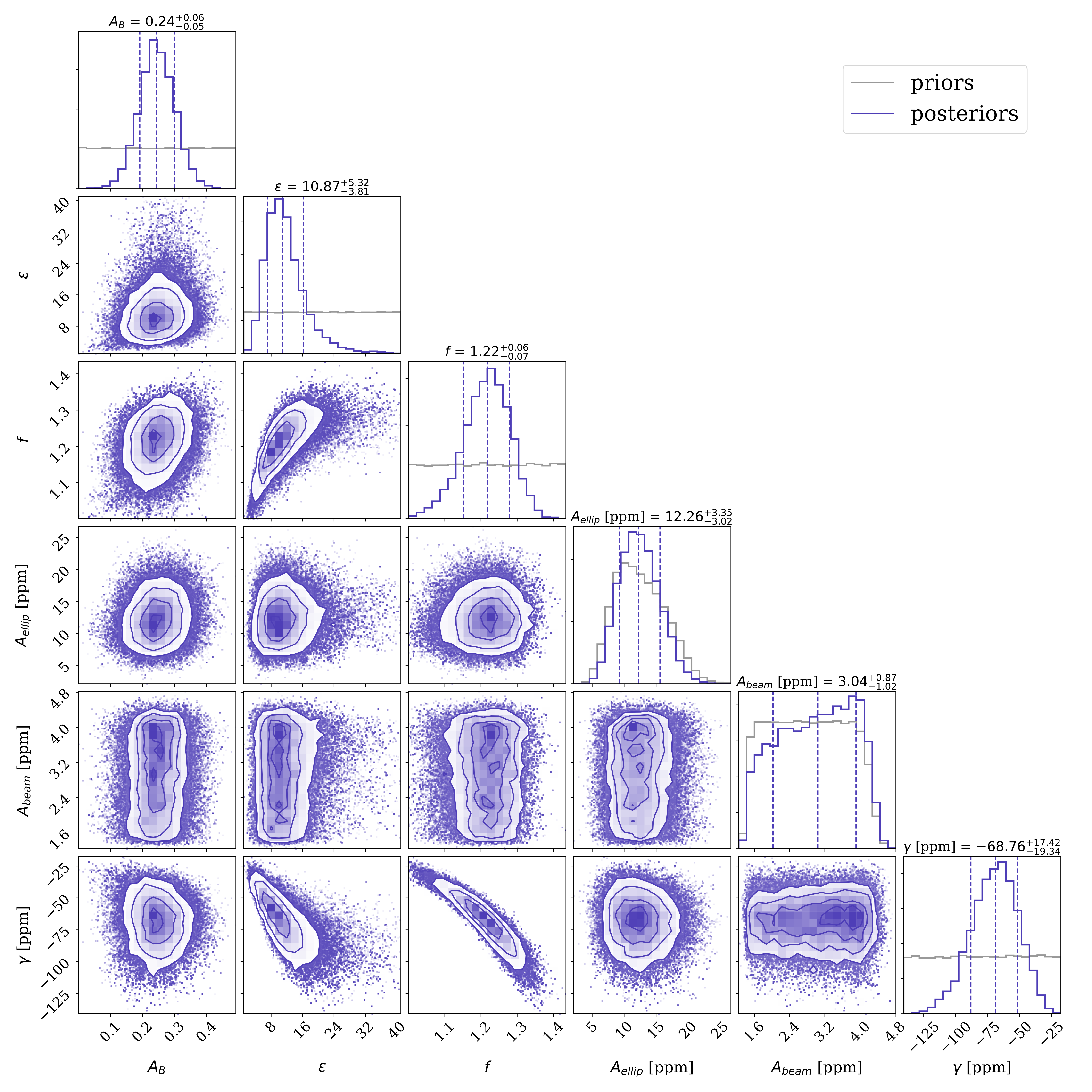}
\caption{
Points and histograms in purple show the posterior distributions for the Bond albedo $A_B$, thermal redistribution efficiency $\epsilon$, intrinsic thermal scaling factor $f$, ellipsoidal variation amplitude $A_{\text{ellip}}$, relativistic beaming amplitude $A_{\text{beam}}$, and vertical offset $\gamma$ in the model fit to the phase curve of the WASP-100 system. Grey histograms show the corresponding prior distributions used in the Bayesian regression analysis.}
\label{fig:corner}
\end{figure*}

The prior distributions for $\left(A\alpha^{-1}\right)_{\mathrm{ellip}}$ and $\left(A\alpha^{-1}\right)_{\mathrm{beam}}$ are then constructed from substituting $10^6$ random samples from the posterior distributions of $K_{rv},\ R_*,\ a,\ P$, and $i$ into Equations~(\ref{eq:Aellip}) \& (\ref{eq:Abeam}). The profiles of all prior distributions discussed in this section can be seen in Figure~\ref{fig:corner}.

\begin{table}
\renewcommand{\arraystretch}{1.25}
\begin{tabular}{lc}
\hline
\multicolumn{2}{l}{Stellar parameters from {\tt isochrones} fits} \\
$T_{eff}$ (K)                          & $6940 \pm 120$                            \\
$R_{\star}$ ($R_{\odot}$)                    & $1.67 ^{+0.18}_{-0.11}$                   \\
g ($\log_{10}\mathrm{cm\ s}^{-2}$) & $4.16^{+0.06}_{-0.08}$\\
$\left[\mathrm{Fe/H}\right]$ & $0.00\pm0.08$\\
$M_{\star}$ ($M_{\odot}$) & $1.47^{+0.06}_{-0.05}$\\
age ($\log_{10}$ yr) & $9.18^{+0.09}_{-0.14}$ \\
$L_{\star}$ ($\log_{10} L_{\odot}$) & $0.76^{+0.09}_{-0.05}$\\
$d$ (pc) & $368.2\pm2.7$\\
$A_V$ & $0.15^{+0.23}_{-0.11}$\\
                                       &                                           \\
                                       
\multicolumn{2}{l}{System parameters from transit fits}                            \\
$P$ (days)                             & $2.849382 \pm 0.000002$                   \\
$\rho_{\star}$ (kg m$^{-3}$)                    & $380_{-13}^{+14}$                              \\
$R_p / R_{\star}$                            & $0.08683 \pm 0.00037$                       \\
$b$ & $0.537_{-0.020}^{+0.017}$\\
$t_0$ (TJD) & $1360.9376\pm0.00001$ \\
$q_1$ & $0.192_{-0.039}^{+0.046}$ \\
$q_2$ & $0.23_{-0.10}^{+0.13}$ \\
                              &                                           \\
\multicolumn{2}{l}{Derived system parameters}                                      \\
$a$ (AU)                               &  $0.043^{+0.005}_{-0.003}$                                         \\
$R_P$ ($R_J$)      &  $1.4^{+0.2}_{-0.1}$                                         \\
$i\ (^{\circ})$ & $84.4\pm0.3$\\
$T_{eq}$ (K)$^{\dagger}$ & $2099\pm38$\\ & \\
From \cite{hellier:2014} \\
$M_p (M_{Jup})$ & $2.03 \pm 0.12$ \\
$K_{rv}$ (km s$^{-1}$) & $0.213 \pm 0.008$ \\
$e$ & 0 ($<0.10$ at $3\sigma$)\\
\hline
\end{tabular}
\caption{System parameters for WASP-100 used in the regression analysis.\newline
$\dagger$ For $A_B=0$}
\label{tab:parameters}
\end{table}

\subsection{Likelihood function}

The likelihood function describes how the data are distributed about the model. A typical approach is to detrend or whiten the data such that the likelihood function is simply a product of Gaussians. In this work, the data has indeed been partially whitened through a processing of photometric detrending. However, the detrending process applied to the phase curve is essentially a low-cut filter (i.e. removes low-frequency correlated noise components), and so leaves frequencies at or above the orbital frequency within the time series. This of course is necessary such that the phase curve signal is not removed, but it means that high frequency stellar noise, such as granulation, could persist in the light curve and would be decidedly non-white. To remedy this, we work with the 500-point binned phase curve for our inference. Because each binned point spans $\sim76$ orbital periods, high frequency noise on top of the orbital frequency will not - in general - be coherent, and thus will average out \citep{pont:2006}. This means that our phase curve data has not only had the low-frequency components suppressed, but the high frequency components too (by a factor of $\sqrt{76} \simeq 8.7$), which justifies our use of a Gaussian likelihood function in what follows.

For the occultation, rather than model the full occultation shape, the key piece of information is the actual depth. Further, since the depth has already been derived using an approach which includes systematic error from detrending differences, we elect to simply include the depth as a single datum in the likelihood function, along with its associated uncertainty. The likelihood function is then

\begin{align}
\log\mathcal{L} = & -\frac{n+1}{2} \log(2\pi) - \sum_{i=1}^n \log\sigma_i - \frac{1}{2}\sum_{i=1}^n\left(\frac{r_i}{\sigma_i}\right)^2 \nonumber\\
\qquad& - \log\sigma_{occ} - \frac{1}{2}\left(\frac{ r_{occ}}{\sigma_{occ}}\right)^2,
\label{eq:likelihood}
\end{align}

where $n$ is the number of real-valued data points, $\sigma_i$ the photometric uncertainty, and $r_i$ the residuals of the phase curve model and the observed data. 

The second line in Equation~(\ref{eq:likelihood}) describes the part of the likelihood function which takes the occultation depth measured in Section \ref{sec:occultation} into consideration when inferring the parameters of the phase curve model, where $\sigma_{occ}$ is the uncertainty of the measured occultation depth and $r_{occ}$ is the difference between a depth sampled randomly from its posterior distribution and the value of the model phase curve at the point in phase where the occultation occurs (i.e. $F(\phi=0)$).

\begin{figure*}
\centering
\includegraphics[width=0.9\textwidth]{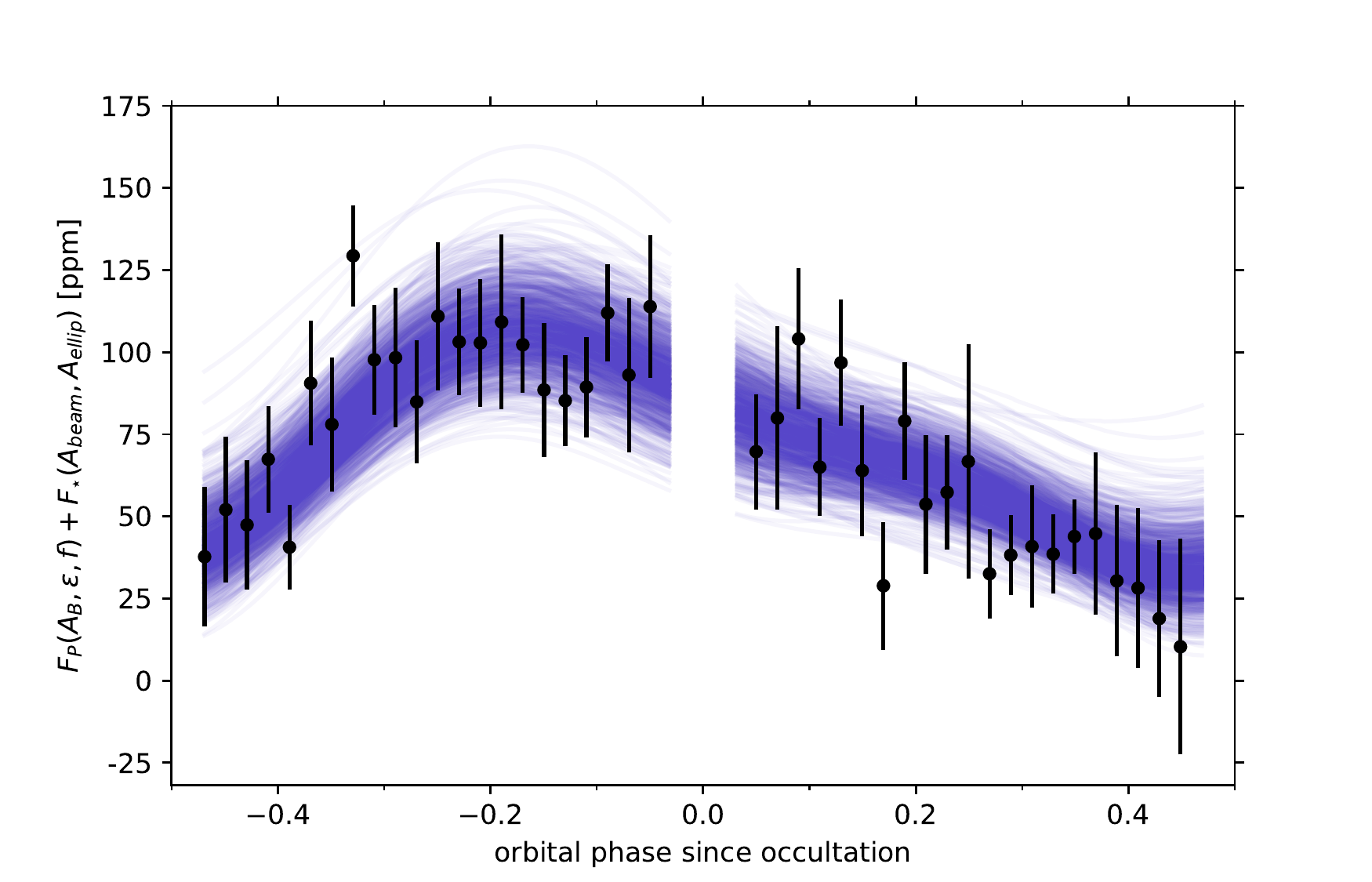}
\caption{
The normalized out-of-transit phase curve of the WASP-100 system averaged across observations made in sectors 1-13 of the \TESS\ data (black points). Error bars are of the $1\sigma$ photometric uncertainties of the data binned into 43 points in phase. The purple lines show a range of phase curve models constructed from 1000 random samples in the posterior distributions from the regression analysis. Blank regions in the phase curve correspond to the occultation (at phase = 0) and transit (phase = $\pm 0.5$), which are masked for analysis of the out-of-transit phase curve. The data have been shifted by the average $\gamma$ value from the posteriors to align with the model samples.
}
\label{fig:phasecurve}
\end{figure*}

\begin{figure*}
\centering
\includegraphics[width=1.0\textwidth]{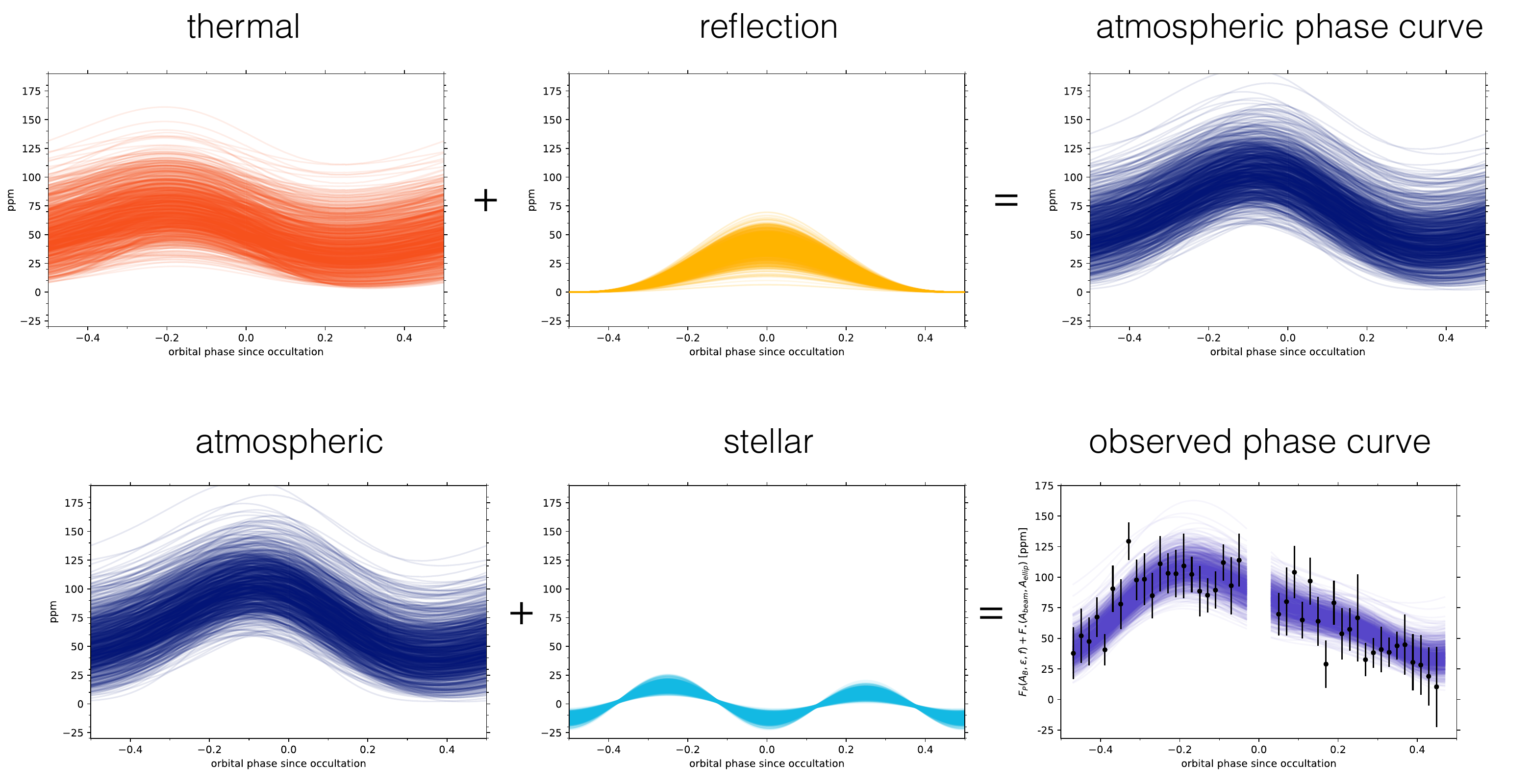}
\caption{Samples from the posterior distribution of the model phase curve of the WASP-100 system deconvolved into the atmospheric components of WASP-100b's phase curve (top) and the separate planet/host components (bottom). The atmospheric component includes both reflected light from the host star and thermal radiation from the planet itself. The phase of maximum brightness in the atmospheric phase curve is offset from where the planet would appear to be in full-phase due to thermal redistribution from the hot day side to the cooler night side, resulting in an eastward deviation of the hotspot from the substellar point.
}
\label{fig:decon_phasecurve}
\end{figure*}

\section{Phase Curve Results}
\label{sec:results}

\subsection{Occultation}

The occultation depth of $(100\pm14\pm16)$\,ppm is remarkably large and it is
interesing to compare this result to theoretical expectation. We do this by
evaluating the limiting case of a blackbody atmosphere (zero albedo) with
no redistribution and completely efficient redistribution. Using the expressions
of \citet{cowanagol:2011}, we propagate our stellar and transit parameter
posterior samples into their expressions for the day-side temperature (Eqn.~4)
and then integrate over the \TESS\ bandpass to predict
$(26_{-3}^{+3} < \delta_{\mathrm{occ}} < 128_{-11}^{+11})$\,ppm. These
extremes correspond to a disk-integrated day-side temperature of $(2098\pm37)$\,K
to $(2680\pm48)$\,K. Accordingly, we conclude that our measured occultation
depth is physically plausible, but towards the upper end of the scale. A
complete interpretation is offered shortly in combination with the phase curve results.

\begin{table}
\renewcommand{\arraystretch}{1.25}
\begin{tabular}{l|r}
Parameter & Value \\
\hline
$A_B$ & $0.24^{+0.06}_{-0.05}$ \\
$A_g^{\dagger}$ & $0.16^{+0.04}_{-0.03}$ \\
$\epsilon$ & $10.9^{+5.3}_{-3.8}$ \\
$f$ & $1.22^{+0.06}_{-0.07}$ \\
$T_{max}$ (K) & $2710\pm100$ \\
$T_{min}$ (K) & $2380^{+170}_{-200}$ \\
$T_{max}$ - $T_{min}$ (K) & $320^{+150}_{-100}$ \\
Thermal hotspot offset ($^\circ$E) & $71^{+2}_{-4}$ \\
Night-side flux at eclipse (ppm) & $50^{+22}_{-19}$ \\
Atmospheric offset ($^\circ$) & $28^{+9}_{-8}$ \\
Atmospheric amplitude$^{\ddagger}$  (ppm) & $62\pm9$\\
Max-brightness offset ($^\circ$) & $63^{+6}_{-8}$ \\
Max-brightness amplitude$^{\ddagger}$  (ppm) & $73\pm9$\\
\hline
\end{tabular}
\caption{Measured and derived values from the model fits to the phase curve of WASP-100b. \newline
$\dagger$ With the assumption of a Lambertian atmosphere such that $A_g = \tfrac{2}{3}A_B$ \newline
$\ddagger$ Amplitude is peak-to-peak} 
\label{tab:results}
\end{table}

\subsection{Non-zero albedo}

From our complete phase curve + occultation model, we show the marginalized
posterior distributions of the model parameters as a corner plot in
Figure~\ref{fig:corner}. Of particular note is that the credible interval for the
Bond albedo is $A_B=(0.24\pm0.06)$, apparently offset away from zero. For a
Lambertian surface, this corresponds to a geometric albedo of
$A_g=(0.16\pm0.04)$. The
marginalized posterior density at $A_B=0$ divided by the prior density yields
the Savage-Dickey ratio \citep{dickey:1971}, an estimate of the Bayes factor
for a nested model. Here, we report a Bayes factor of 165 in very strong favor of a
non-zero albedo.

\subsection{Warm night side}

The ratio between the radiative timescale and advective timescale of the atmospheric height probed by \TESS\ is measured to be $\epsilon = 10.9^{+5.3}_{-3.8}$, indicating heat transport from the substellar point to the nightside in an eastward direction, i.e. the same direction as the rotation of the surface assuming WASP-100b is on a prograde orbit. This redistribution of heat causes an eastward shift of the brightest region in the atmosphere of WASP-100b. Using the Savage-Dickey ratio to test the $\epsilon = 0$ case in which there is no thermal redistribution returns a Bayes factor of 151, which is in very strong favor of the model with efficient thermal redistribution. This result is supported by the fact that the occultation depth exceeds the peak-to-peak phase curve amplitude, implying a hot night side. It will be important to verify this result using other observatories; additionally \TESS\ Cycle 3, which will revisit the field hosting WASP-100.

\subsection{Evidence for winds}

We find that the phase of maximum brightness occurs $(63^{+6}_{-8})^{\circ}$ prior to the phase of occultation. However, the phase of maximum brightness in the observed phase curve seen in Figure~\ref{fig:phasecurve} does not correlate to the phase shift of the atmospheric signal, since the observed phase curve is a convolution of the atmospheric phase curve and the coherent photometric effects of the star. This is illustrated in the bottom panel of Figure \ref{fig:decon_phasecurve}, which shows the observed phase curve deconvolved into its stellar and atmospheric components according to samples from the regression analysis.

Additionally, because we measure a significant contribution of reflected light to the atmospheric phase curve, even the phase shift of the atmospheric component, measured to be $(28^{+9}_{-8})^{\circ}$, does not directly correlate to the offset of the hottest longitude from the substellar point (see the top panel of Figure~\ref{fig:decon_phasecurve}). After deconvolving the thermal and reflected components of WASP-100b's phase curve, we measure a longitudinal hotspot offset of $(71^{+2}_{-4})^{\circ}$ east of the substellar point. We measure the brightness temperature of the hottest spot to be $(2720\pm150)$\,K, where the temperature of the coolest spot is $(2400\pm220)$\,K, giving a longitudinal temperature contrast of $\Delta T=(320^{+150}_{-100})$\,K.

As a point of reference, we calculate the theoretical maximum peak temperature of WASP-100b using the expression for the temperature at the substellar point from \citet{cowan:2011} (Eqn. 4), assuming a zero albedo and no thermal redistribution. Leveraging our posterior samples from the transit and isochrone solution, we estimate that the expected maximum peak temperature should be no more than $T=(2967\pm53)$\,K, which the credible interval of our maximum measured temperature is indeed below.

\subsection{Additional heating?}

The intrinsic thermal scaling factor $f$, which signifies a deviation from WASP-100b's equilibrium temperature, is measured to be $f = 1.22\pm0.07$. This suggests some modest evidence for additional heating from an internal source, but with a Bayes factor of only 3.9 compared to the $f=1$ case, we caution that this result is somewhat marginal and cannot be verified using the independent occultation measurement on its own. Further, the theoretical maximum
temperature is clearly above our inferred peak temperature, suggesting that no extra heating is needed. However, that calculation
assumes zero albedo and our models favor a modest albedo which explains the behavior of the $f$ factor inflating slightly above unity to compensate for that energy loss.

It is thus apparent that a slight degeneracy exists between a moderate-$\epsilon$/moderate-$f$ model and a greater-$\epsilon$/greater-$f$ model, which can be explained by the change in the phase curve amplitude with $\epsilon$ and the flexible nature of the model's vertical baseline. When the thermal redistribution efficiency of the atmosphere is high, the air mass heated at the substellar longitude is redistributed toward the nonirradiated hemisphere in an attempt to reach thermodynamic equilibrium, which results in a phase curve with a phase-shifted ``hot-spot,'' and a diminished thermal amplitude due to the decrease in temperature contrast between the day and night sides. Consequently, the vertical baseline of the thermal phase curve will be greater for a more thermally redistributed atmosphere because the nightside has a higher temperature than one in which there is little to no thermal redistribution.

A model which fits well to the phase offset and amplitude of the data in Figure \ref{fig:phasecurve} can be constructed with a high thermal redistribution efficiency $\epsilon$ and intrinsic thermal scalar $f$. However, the data can also be well represented by an even larger $\epsilon$ as long as $f$, and therefore the average temperature of WASP-100b, is also increased in order to maintain the thermal phase curve amplitude that is given by the out-of-transit data and measured occultation depth. When $\epsilon$ and $f$ increase, the vertical baseline of the model shifts upward, and therefore the magnitude of the baseline correction also increases to minimize $\chi^2$. Once $\epsilon \gg 1$, further increasing the magnitude of $\epsilon$ has little affect on the phase offset of the point of maximum brightness.

\subsection{Gravitational effects}

We are unable to measure the Doppler beaming amplitude of WASP-100, though this is to be expected for an upper limit of $<4.2$ ppm at 95\% confidence on the amplitude of this signal. The data do not appear to put a significantly tighter constraint on the ellipsoidal variation amplitude compared to its prior distribution (Figure \ref{fig:corner}), which has an amplitude distribution of $(12\pm3)$\,ppm. The posterior distribution of the ellipsoidal variation amplitude corresponds to a planetary mass of $1.97^{+0.48}_{-0.40}\,M_{\jup}$, which is in agreement with the prior set by the mass measured by the radial velocity observations in \cite{hellier:2014}.

\subsection{Statistical significance}

To gauge the statistical significance of our results, we conduct additional regression analyses for five simpler phase curve models and compare their Bayesian information criteria (BIC, \citealt{schwarz:1978}) and Akaike information criteria (AIC, \citealt{akaike:1974}) to that of the full model which describes the atmospheric phase curve and  the stellar contribution to the phase curve (Equation \ref{eq:forwardmodel}). Both criteria measure model likelihood while penalizing a higher number of free parameters.

The simplest model is that of the null result, which in this case is a flat line constant in phase. When comparing the likelihood of the fit to the full model to that of the null model, we compute $\Delta\text{BIC} = 25$ in strong favor of the full phase curve model. The AIC is even less punitive toward the number of free parameters, for which we measure $\Delta\text{AIC}= 53$ in very strong favor of the full model.

To judge whether the thermal and reflection components of the atmospheric phase curve are significantly retrieved, we repeat our regression analysis against models which exclude these atmospheric components, i.e. Equation~\ref{eq:forwardmodel} in the case that i) $F_P = F_R(\phi, A_B)$ and ii) $F_P = F_T(\phi, A_B, f, \epsilon)$. For the case in which the atmospheric phase curve is modeled by only a thermal component, $\Delta\text{BIC}=6$ in strong favor of the full model. The case in which the atmospheric phase curve is modeled only by reflection performs even more poorly, measuring $\Delta\text{BIC}=13$ in even stronger favor for the model which includes both atmospheric components.

We then perform additional regression analyses on a pair of models which do not include the stellar components of the phase curve, namely the Doppler beaming and ellipsoidal variation effects. The model excluding ellipsoidal variations is strongly preferred over the full model, with $\Delta\text{BIC}=6$. For the case in which $A_{\mathrm{beam}} = 0$, $\Delta\text{BIC}=6$ compared to the model without ellipsoidal variations, and $\Delta\text{BIC}>10$ against all of the other models, which makes the phase curve model without Doppler beaming the most significantly preferred model of the six tested here.

From the statistical analysis discussed in this section, we conclude that the phase curve signal of WASP-100b is real and significant. The values for the BIC and AIC for each of the models tested in this section can be seen in Table~\ref{tab:modelstats}.

\begin{table}
\caption{Model selection statistics on the phase curve of WASP-100b (binned into 439 points). The ``full model'' refers to Equation~\ref{eq:forwardmodel}, while models labeled ``full model $-$ [component]'' refer to model scenarios which do not include said component. Values correspond to the median models constructed from 1000 samples from their respective posterior distributions.\label{tab:modelstats}}
\begin{tabular}{l|ccc|ccc}
\hline
\hline
Models & $\chi^2$ & BIC & AIC \\
\hline
Flat line & 452 & 458 & 454 \\
Full model & 385 & 434 & 401\\
Full model $-$ beaming & 385 & 421 & 397\\
Full model $-$ ellipsoidal & 391 & 428 & 403\\
Full model $-$ reflection & 390 & 439 & 406 \\
Full model $-$ thermal & 409 & 446 & 421 \\
\end{tabular}

\begin{tabular}{lcc}\hline\hline
\multicolumn{1}{l|}{Flat line model against the...} & \multicolumn{1}{c|}{$\Delta$BIC} & \multicolumn{1}{c|}{$\Delta$AIC} \\ \hline
\multicolumn{1}{l|}{full model} & \multicolumn{1}{c|}{24.7} & \multicolumn{1}{c|}{53.2}\\
\multicolumn{1}{l|}{full model $-$ beaming} & \multicolumn{1}{c|}{36.9} & \multicolumn{1}{c|}{57.3}  \\
\multicolumn{1}{l|}{full model $-$ ellipsoidal} & \multicolumn{1}{c|}{30.8} & \multicolumn{1}{c|}{51.2}  \\
\multicolumn{1}{l|}{full model $-$ reflection} & \multicolumn{1}{c|}{19.8} & \multicolumn{1}{c|}{48.4}\\
\multicolumn{1}{l|}{full model $-$ thermal} & \multicolumn{1}{c|}{12.8} & \multicolumn{1}{c|}{33.2}\\
\end{tabular}
\end{table}

\subsection{Addressing the effect of \TESS's momentum dumps and other aspects of data reduction}

The reaction wheels on the \TESS\ spacecraft experience a build up of momentum which is corrected for by resetting the reaction wheel speeds to lower values approximately once every 2.5 days, where each momentum dump causes a momentary increase in the spacecraft's pointing instability\footnote{See the \hyperlink{https://archive.stsci.edu/tess/tess_drn.html}{\TESS\ Data Release Notes}}. The occurrence rate of these momentum dumps is close enough to the orbital period of the planet ($\sim 2.8$ days) to elicit some concern for the potential effect this may have on WASP-100b's phase curve. 

To measure the magnitude of this effect, we construct a model of the momentum dump profile in phase with WASP-100b and measure its maximum peak-to-peak amplitude. To do this, we first locate the time of each momentum dump in all sectors of WASP-100's observation and fold the detrended light curve as a function of time since the momentum dump. We then construct a model for the momentum dump profile by fitting a suite of polynomial functions of 0$^{\mathrm{th}}$ to $20^{\mathrm{th}}$ order using weighted linear least squares, and select the polynomial which produces the lowest Akaike Information Criterion. We then unfold the noise-less polynomial model back into a function of time, and refold into phase with WASP-100b. The resulting profile has a maximum peak-to-peak amplitude significantly less than the peak-to-peak amplitude of the phase curve of WASP-100b, and is less than its $2\sigma$ error ($74 \pm 10~\mathrm{ppm}$). This analysis was repeated for both the \phasma\ detrended light curve and the \texttt{slowpoly} detrended light curve, and for momentum dump models chosen by the Bayesian Information Criterion, each case showing the same result. From this we conclude that the momentum dumps of the spacecraft's reaction wheels have an insignificant effect on the phase curve of WASP-100b.

Additionally, because each sector of observation comes with its own anomalies, we examine the effect each sector has on the binned phase curve of WASP-100b by removing one sector from the time series and comparing the resulting phase curve to the full 13-sector phase curve used in our analysis. In each case, the binned data are all well within the 2$\sigma$ error of the binned data in the full phase curve, indicating that no one sector is significantly affecting the profile of the phase curve of WASP-100b.

To examine if our choice of binning statistic has an effect on the results presented in the previous sections, we repeat the regression analysis for the phase curve constructed with median binning, and find that all results presented in Table~\ref{tab:results} agree within $1\sigma$ significance.

\section{Discussion}
\label{sec:discussion}

The results presented in this paper suggest that the atmosphere of WASP-100b is likely to have a strong thermal redistribution efficiency indicative of atmospheric winds, with significant reflectivity in the \TESS\ waveband. From the measured occultation depth and regression analysis of the phase curve, we measure a maximum dayside temperature of $2720\pm150$ K, placing WASP-100b in the ``ultra hot'' class of Jupiter-sized exoplanets \citep{parmentier:2018b, arcangeli:2018, bell:2018}. Our study provides more insight into this relatively new class of exoplanets.

\subsection{WASP-100b in context}

Of the three hot-Jupiter phase curves which have been observed by \TESS\ so far (\citealt{shporer:2019, daylan:2019, bourrier:2019, wong:2019}), WASP-100b is the first to show a thermal phase shift indicative of efficient heat transport in its atmosphere. The magnitude of its $71\pm4^{\circ}$ hotspot offset is rivaled only by the phase shift of $\upsilon$ Andromedae b, which has been measured to be $(84.5 \pm 2.3)^{\circ}$ \citep{crossfield:2010}. Such a large thermal phase shift is unexpected for ultra-hot Jupiters such as WASP-100b, which have been predicted to have much shorter radiative time scales than thermal redistribution time scales, and therefore negligible phase curve offsets \citep{perez:2013, komacek:2016, komacek:2017, schwartz:2017}. The eastward direction of WASP-100b's hotspot offset is, however, typical of hot-Jupiters which have been previously observed to have asymmetric thermal phase curves (\citealt{parmentier:2018a} and references therein).

\cite{bell:2018} have recently suggested a possible mechanism for increased heat transport specific to ultra-hot Jupiters, which may help explain WASP-100b's unexpectedly large eastward hotspot offset. On the day sides of these worlds, temperatures are hot enough to dissociate hydrogen molecules. This hydrogen gas is then carried by eastward winds from the sub-stellar point to cooler longitudes, where temperatures are low enough to allow for the recombination of H$_2$. This recombination is a highly exothermic process, releasing a significant amount of energy via latent heat to the surrounding gas. On the opposite hemisphere, the recombined molecular gas is carried back to the hotter day side, where latent heat is used in the redissociation of H$_2$, effectively cooling the longitudes west of the sub-stellar point.

Indeed, when \cite{komacek:2018} included heat transport via H$_2$ dissociation/recombination in a follow-up study to \cite{komacek:2016}, they found that the day-night temperature contrast of ultra-hot Jupiters decreased with increasing incident stellar flux, the opposite conclusion of their previous theoretical analysis. A quantitative analysis of the significance of H$_2$ dissociation/recombination in the heat recirculation of WASP-100b is a bit beyond the scope of this study, although we acknowledge it would certainly be worth exploring.

Our ability to measure such a large shift in the phase curve of WASP-100b may be a product of the long observational baseline of the WASP-100 system. A majority of the planets viewed by \TESS\ can only be observed for a maximum of 27 days, i.e. the duration of one observational sector in the \TESS\ mission. For planets with peak-to-peak phase curve amplitudes as low as WASP-100b's (Table~\ref{tab:results}), the majority will not have enough data to detect a phase curve with a significant signal to noise ratio. The convenient location of WASP-100 in the continuous viewing zone of \TESS\ has allowed its observation through many orbits, therefore strengthening the signal to noise ratio of a phase curve amplitude which is relatively small due to the decreased longitudinal temperature contrast of WASP-100b.

WASP-100b's geometric albedo of $0.17\pm0.05$ is among the highest measured from a \TESS\ phase curve so far, comparable to the geometric albedo of WASP-19b \citep{wong:2019} and significantly greater than that measured of WASP-18b \citep{shporer:2019} and WASP-121b \citep{daylan:2019}. The albedo we measure is in line with the expectation for hot-Jupiters to have a relatively low reflectivity in the optical to near-infrared transition regime \citep{mallonn:2019} and is similar to that of several hot-Jupiters observed by \Kepler\ \citep{angerhausen:2015}.

\subsection{Caveats}

In modeling the phase curve of a star-planet system, there is some danger in confusing a star's relativistic beaming signal for an eastward offset of the hotspot in the planet's atmospheric phase curve. Fortunately we are able to break this degeneracy for the WASP-100 system with the radial velocity measurements presented in \cite{hellier:2014} (see Table~\ref{tab:parameters}), which we use to construct an informative prior on the magnitude of the star's relativistic beaming amplitude in our regression analysis. 

The main limitations of our atmospheric phase curve model lie in are our assumptions that the atmosphere of WASP-100b is Lambertian, and radiates as a blackbody. If the atmosphere of WASP-100b is in actuality composed of many particles that scatter photons in a preferential direction, the relative contributions of the modeled reflection and thermal components to the observed phase curve would have to be altered accordingly \citep{dyudina:2005}. Approximating WASP-100b's spectrum as that of a blackbody could be an adequate model for the dayside spectrum, where temperatures are high enough to dissociate absorbent molecules such as H$_2$O, TiO, and VO, and to support a H$^-$ continuum opacity \citep{arcangeli:2018, lothringer:2018}. Because we measure a high redistribution efficiency of heat from the day side to the night side on WASP-100b, the $2400\pm200$ K nightside temperature we derive from the model fit may also be high enough to justify the blackbody approximation, although the posterior distribution of this measurement indicates the nightside temperature is likely low enough to maintain the molecular bond of prominent visible-infrared absorbers such as titanium oxide and vanadium oxide \citep{lothringer:2018}, in which case the blackbody assumption would not hold.

\section*{Acknowledgments}

This paper includes data collected by the \TESS\ mission, which are publicly available from the Mikulski Archive for Space Telescopes (MAST). Funding for the TESS mission is provided by the NASA Explorer Program. This research has made use of the {\tt corner.py} code and {\tt emcee} package by Dan Foreman-Mackey at
\href{http://github.com/dfm/corner.py}{github.com/dfm/corner.py} and \href{https://github.com/dfm/emcee}{github.com/dfm/emcee}, {\tt scipy}, {\tt astropy}, and the NASA
Exoplanet Archive, which is operated by the California Institute of Technology,
under contract with the National Aeronautics and Space Administration under the
Exoplanet Exploration Program.
We would also like to thank Ian Wong and Avi Shporer for their helpful correspondence and insight.

\appendix

\section{non-parametric detrending with {\tt phasma}}\label{sec:appendix}

In \cite{jansen:2018}, we state that the final phase curve function we want can be expressed as
\begin{align}\label{eq:phasecurve}
    \widetilde{F}(t) &= \frac{F_P(t)}{\overline{F_{\star}}}\nonumber\\
                  &= \frac{F(t) - G(t)}{G(t)}
\end{align}
where $F(t)$ is the observed light curve, i.e. $F(t) = F_P(t) + F_{\star}(t)$, and $G(t)$ is the ``nuisance'' function defined by 

\begin{equation*}
    G(t) = \overline{F_P} + \Pi(t)*F_{\star}(t)
\end{equation*}
where
\begin{equation*}
\Pi(t) =
    \begin{cases}
   1 & \text{for } t - \frac{P}{2} < t < t + \frac{P}{2}\\    
   0 & \text{else}    
\end{cases}.
\end{equation*}
In other words, $G(t)$ is a moving mean function with window = $P$. The stellar flux $F_{\star}(t)$ can also be expressed by a sum of some low frequency oscillations (where we define ``low'' to mean frequencies much less than the planet's orbital frequency) due to e.g. spots on the surface of a slow rotator, and much higher frequency residuals about this low frequency mean:
\begin{equation}\label{eq:stellarflux}
    F_{\star}(t) = F_{\star,low}(t) + \Delta F_{\star}(t).
\end{equation}
Our first assumption is that the residuals about the lower frequency stellar signal are approximately equally mixed between positive and negative values, in which case their sum about the mean converges on zero. Therefore $G(t)$ is reduced to \begin{equation}
    G(t) = \overline{F_P} + F_{\star,low}(t)
\end{equation}
Substituting the above back into Equation \ref{eq:phasecurve}, we get
\begin{align}\label{eq:phasecurve2}
    \widetilde{F}(t) &= \frac{F_P(t) + F_{\star}(t) - [\overline{F_P} + F_{\star,low}(t)]}{\overline{F_P} + F_{\star,low}(t)}\nonumber\\
    &= \frac{F_P(t) + F_{\star}(t)}{\overline{F_P} + F_{\star,low}(t)}- 1.
\end{align}
If we make a second assumption that $\overline{F_P} << F_{\star,low}(t)$, Equation \ref{eq:phasecurve2} becomes
\begin{equation}
    \widetilde{F}(t) = \frac{F_P(t)}{ F_{\star,low}(t)} + \frac{ F_{\star,low}(t) + \Delta F_{\star}(t)}{ F_{\star,low}(t)} - 1.
\end{equation}
Folding $\widetilde{F}(t)$ about the exact period of the planet, the average of $\Delta F_{\star}(t)$ approaches zero as a consequence of our first assumption (although we account for any residual noise as a constant nonetheless), and $F_{\star,low}(t)$ averages to $\overline{F}_{\star, low}$. If $\Delta F_{\star}(t)$ averages to zero upon a phase fold, then from Equation  \ref{eq:stellarflux} it should be true that $\overline{F}_{\star, low} \approx \overline{F_{\star}}$. Thus we get
\begin{align}
    \text{fold}_P(\widetilde{F}(t)) &= \text{fold}_P\left(\frac{F_P(t)}{ F_{\star,low}(t)} + \frac{ F_{\star,low}(t) + \Delta F_{\star}(t)}{ F_{\star,low}(t)} - 1\right)\nonumber\\
    &\approx \text{fold}_P\left(\frac{F_P(t)}{ F_{\star,low}(t)}\right) + \frac{\overline{F}_{\star, low}}{\overline{F}_{\star, low}} +  \gamma - 1 \nonumber\\
    &\approx \frac{F_P(\phi)}{\overline{F_{\star}}} + \gamma,
\end{align}
the desired function plus a possible vertical offset $\gamma$, a constant which is left as a free parameter in our model fitting.

\bsp
\label{lastpage}
\end{document}